\documentclass{scrartcl}

\usepackage[utf8]{inputenc}
\usepackage[T1]{fontenc}
\usepackage[margin=12mm]{geometry}
\usepackage[svgnames]{xcolor}
\usepackage{multicol,enumitem,microtype,lmodern}
\setlist{leftmargin=*}

\usepackage{mathtools,amssymb,tensor,siunitx}

\usepackage[colorlinks=true,allcolors=DarkBlue]{hyperref}
\usepackage{cleveref}
\crefname{section}{sec.}{secs.}
\crefname{appendix}{app.}{apps.}
\Crefname{equation}{Eq.}{Eqs.}

\usepackage[backend=biber,sorting=none,citestyle=numeric-comp,giveninits=true,doi=false,date=year,url=false]{biblatex}
\renewbibmacro{in:}{}

\AtEveryBibitem{\clearfield{number}\clearfield{issue}\clearfield{url}}
\DeclareFieldFormat[article]{volume}{\mkbibbold{#1}}
\DeclareFieldFormat{eprint:arxiv}{\href{http://arxiv.org/abs/#1}{#1}}
\DeclareFieldFormat{pages}{\mkfirstpage{#1}}

\newcommand{\dif}{\mathrm{d}}

\newenvironment{alignedeqn}{\begin{equation}\begin{aligned}}{\end{aligned}\end{equation}\ignorespacesafterend}

\begin{document}

\title{Cosmology and the Fate of Dilatation Symmetry}
\author{C. Wetterich\\
	\small\href{mailto:c.wetterich@thphys.uni-heidelberg.de}{c.wetterich@thphys.uni-heidelberg.de}\\
	\small Deutsches Elektronen Synchrotron DESY, Hamburg, Germany}
\date{\small 24 September 1987, published in \href{http://www.sciencedirect.com/science/article/pii/0550321388901939}{Nuclear Physics B 302 (1988) 668}}
\maketitle

\begin{abstract}
	We discuss the cosmological constant problem in the light of dilatation symmetry and its possible anomaly. For dilatation symmetric quantum theories realistic asymptotic cosmology is obtained provided the effective potential has a non-trivial minimum. For theories with dilatation anomaly one needs as a non-trivial ``cosmon condition'' that the energy-momentum tensor in the vacuum is purely anomalous. Such a condition is related to the short-distance renormalization group behavior of the fundamental theory. Observable deviations from the standard hot big bang cosmology are possible.
\end{abstract}

\BeforeStartingTOC[toc]{\begin{multicols}{2}}
\AfterStartingTOC[toc]{\end{multicols}}
{\setlength{\columnseprule}{0.4pt}
\setlength{\columnsep}{1cm}
\hypersetup{hidelinks}\tableofcontents}
\bigskip

\begin{multicols}{2}

\section{Introduction}

Theories with only dimensionless parameters are described by a dilatation-invariant (classical) action. This symmetry consists of a common multiplicative scaling of all fields according to their dimension. The ``fundamental constants'' with dimension of mass, like the electron mass $m_e$ and the Planck mass $M_\text{P}$, are typically induced by vacuum expectation values (VEVs) of scalar fields. In general, such VEVs may vary as a consequence of cosmological evolution and the observed values of the corresponding ``fundamental constants'' are obtained only as a result of asymptotic ``late'' cosmology. Quantum fluctuations may or may not conserve dilatation symmetry. In the second case dilatation symmetry is anomalous and an intrinsic scale $m$ is introduced by quantization ($m$ is proportional to the renormalization scale $\mu$).

In ref.~\cite{wetterich1988cosmologies} we have studied cosmologies with a variable Newton's ``constant''. In these models $M_\text{P}$ is generated by the VEV of a scalar singlet $\chi$ and $m_e$ is proportional to the Higgs doublet $\tilde\varphi$ of the standard model. As an approximation to the effective action we used a Brans-Dicke-type theory \cite[eq.~(2.5)]{wetterich1988cosmologies}
\begin{alignedeqn}\label{eqn:brans-dicke action}
	S^0
	= -\int \dif^4 x \, \tilde{g}^{1/2} \Bigl\{&\chi^2 \tilde{R} - 4 \omega \partial_\mu \chi \partial^\mu \chi\\
	&- \partial_\mu \tilde\varphi \partial^\mu \tilde\varphi + V(\tilde\varphi,\chi)\Bigr\}.
\end{alignedeqn}
The dynamics of these models depends critically on the form of the effective potential $V$. In particular, all effects from anomalies will appear through $V$. (For a more general situation see \cref{sec:dilatation anomalies,sec:appendix}.)

Consider first the case where the fundamental quantum theory does not lead to any intrinsic mass scale $m$ (dilatation symmetry is anomaly-free). Then physics can only depend on \textit{scale ratios} like $\tilde\varphi/\chi$, but not on $\tilde\varphi$ and $\chi$ separately. In particular the effective potential must have the form $V = \tilde\varphi^4 v(\tilde\varphi/\chi)$. The theory has a global dilatation symmetry corresponding to a constant scaling of all fields according to their dimension. (In this context the inverse metric $\tilde{g}^{\mu\nu}$ has the same scale dimension as $\chi^2$.) Since we observe the appearance of scales in our world, dilatation symmetry must be spontaneously broken. Any non-zero VEV of a scalar field induces such a spontaneous breaking. The scale characteristic for spontaneous dilatation symmetry breaking may be identified with the largest VEV of a scalar. In our case it is given by $\chi$ and should be in the vicinity of $M_\text{P}$. A spontaneously broken global symmetry leads to a Goldstone boson, the dilaton, which should only have derivative couplings. These couplings are suppressed by powers of $M_\text{P}^{-1}$. A shift in the dilaton field corresponds to an overall change of all scales. In a theory where only scale ratios are measurable the overall scale plays the same unobservable role as the phase in a theory with global $U(1)$ symmetry. In our model the dilaton can be identified with the field $\sigma \sim \ln \chi$.

A fundamental quantum theory without intrinsic scale $m$ should be finite. As an alternative one may consider an asymptotically free renormalizable theory which has a running dimensionless coupling constant. Even though the classical action may not have any scale parameter, a renormalization scale $\mu$ must be introduced in the quantization procedure. This leads to the appearance of an intrinsic mass scale $m$ (which plays the same role as $\Lambda_\text{QCD}$ in a pure QCD theory). Dilatation symmetry is said to have anomalies -- it is not realized as a quantum symmetry. Nevertheless, for $m$ much smaller than the scale $\chi$ characteristic for spontaneous dilatation symmetry breaking, we can still consider the dilatation symmetric theory as an approximation. The language of symmetry currents etc.\ remains useful, but the anomaly leads to some characteristic qualitative changes. The physical quantities are no longer independent of the dilaton VEV since the overall scale ``feels'' the existence of an intrinsic scale $m$, even if the connection is only weak. As a consequence, the dilaton has not only derivative couplings. It is subject to a driving force proportional to the dilatation anomaly, which is given by the anomalous trace of the energy-momentum tensor $\tensor{\tilde\vartheta}{_\mu^\mu}$. It also acquires a small mass, typically suppressed by powers of $m/\chi$. Any vacuum solution with static constant $\chi$ requires the anomaly to vanish. In general the anomaly depends on $\chi$. This governs the dynamic behavior for $\chi$. As a consequence, the dilatation anomaly determines those qualitative properties of the effective potential which characterize the asymptotic evolution of cosmology. Our study of cosmologies with dynamical Planck mass in \cite{wetterich1988cosmologies} is therefore intimately connected with the fate of dilatation symmetry. In this sense the present paper should be understood as a logical continuation of \cite{wetterich1988cosmologies}. Although our treatment of dilatation symmetry is essentially self-contained, we recommend reading ref.~\cite{wetterich1988cosmologies} for a more profound understanding of the spirit, formalism and notations of the present work.

In \cref{sec:models without mass scales} we study the cosmology of scale-free models (without dilatation anomaly). We find that late cosmology leads to the standard big-bang picture provided the potential $V(\tilde\varphi,\chi)$ has a non-trivial minimum. In this case the dilaton mode becomes irrelevant for late cosmology (in the limit where its coupling to matter can be neglected). On the other hand, if $V(\tilde\varphi,\chi)$ has only a relative minimum with respect to $\tilde\varphi$ the cosmology looks like the standard model with non-vanishing cosmological constant. In \cref{sec:dilatation anomalies} we turn to models with dilatation anomalies. We formulate three conditions on the dynamics of the dilaton which are necessary for a realistic cosmology. The trace anomaly should vanish for some value of the dilaton field, $\tensor{\tilde\vartheta}{_\mu^\mu}(\sigma_0) = 0$. For this value the dilaton mass should be positive. Finally, for the static vacuum solution with $\sigma = \sigma_0$ the trace of the energy-momentum tensor should be purely anomalous. If the dilaton fulfills these three conditions it is called a cosmon \cite{peccei1987adjusting}. Its dynamics drives the cosmological constant to zero. In \cref{sec:cosmon condition} we establish the connection between the ``cosmon condition'' and the short-distance behavior of the underlying fundamental theory for models where intrinsic mass scales arise only from the running of dimensionless couplings. One finds that the trace anomaly for static configurations is given by the renormalization group equation for the effective potential, $\tensor{\tilde\vartheta}{_\mu^\mu} = \mu \, \partial V/\partial \mu$. We discuss in \cref{sec:anomalous renormalization group equation} the situation where this renormalization group equation is governed by an anomalous dimension, $\mu \, \partial V/\partial \mu = A \, V$. The cosmology for this case is investigated in \cref{sec:cosmology with time variation}. It
is characterized by a cosmological ``constant'' which evolves with time and is of the general type discussed in ref.~\cite{reuter1987time}. Realistic cosmology is obtained if the anomalous dimension $A$ is within a certain range. We conclude this paper in \cref{sec:conclusions} with a discussion of the cosmological constant problem.

\section{Models without mass scales}
\label{sec:models without mass scales}

Let us first investigate models without intrinsic mass scale. The most general form\footnote{The discussion of this section formally includes terms like $\tilde\varphi^2 \, \chi^2$ or $\chi^4$. They are, however, at variance to the spirit of \cite{wetterich1988cosmologies}.} for the effective potential is
\begin{equation}\label{eqn:effective potential}
	V(\tilde\varphi,\chi)
	= v(\tilde\varphi/\chi) \, \tilde\varphi^4
	= v(x) \tilde\varphi^4,
\end{equation}
with $v$ a (dimensionless) function depending only on the ratio $x = \tilde\varphi/\chi$. Possible extrema of $V$ with respect to $\tilde\varphi$ are determined by
\begin{equation}\label{eqn:v extrema}
	\frac{\partial V}{\partial \tilde\varphi}
	= \biggl(4 v(x) + x \, \frac{\partial v}{\partial x}\biggr) \tilde\varphi^3
	= 0.
\end{equation}
If \cref{eqn:v extrema} has a solution with $\tilde\varphi \neq 0$, the corresponding value of $\tilde\varphi$ must be proportional to $\chi$. We therefore expect the existence of cosmologies where any change in $\chi$ is accompanied by an appropriate change in $\tilde\varphi$ so that $\tilde\varphi/\chi$ remains constant.

This can be seen more easily by performing a Weyl scaling of the metric \cite[sec.~4]{wetterich1988cosmologies}. The rescaled potential now reads ($\sigma = M \ln(\chi/M)$, $\varphi = \tilde\varphi \, M/\chi$)
\begin{alignedeqn}
	W(\varphi,\sigma)
	&= v(\varphi/M) \, \varphi^4
	= v(x) \, x^4 \, M^4,\\
	x
	&= \frac{\tilde\varphi}{\chi}
	= \frac{\varphi}{M}.
\end{alignedeqn}
We note that $W$ is completely independent of $\sigma$. Therefore $\sigma$ is a massless Goldstone boson which has only derivative couplings (compare \cite[eq.~(4.6)]{wetterich1988cosmologies}). All of particle physics depends only on the ratio $x = \varphi/M$. The scale $M$ itself is arbitrary and one obtains equivalent physics for any choice of $M$. For ``quasistatic'' cosmologies with constant Hubble parameter $\tilde{H}$ one out of the three scales $\tilde\varphi$, $\chi$ and $\tilde{H}$ is irrelevant. Physics depends only on the ratios $\tilde\varphi/\chi$ and $\tilde{H}/\chi$. This generalizes to evolutionary cosmologies which depend in addition on ratios like $\dot{\tilde{H}}/\chi^2$ etc.

The asymptotic behavior of ``scale-free'' cosmologies depends on the possible existence of a minimum of $W$ which must obey\footnote{We disregard here the uninteresting trivial case of a minimum at $\varphi = 0$.}
\begin{equation}\label{eqn:min cond}
	x_0 \, \frac{\partial v}{\partial x}(x_0)
	= -4 v(x_0).
\end{equation}
If \cref{eqn:min cond} has no solution, the field $\varphi$ cannot be asymptotically static and one is confronted with the problems of cosmologies with varying $\varphi^2 G_\text{N}$ (similar to the case $\varphi \sim t^\alpha$ described in \cite[sec.~5]{wetterich1988cosmologies}). On the other hand, if a non-trivial minimum of $W$ exists, it is reasonable to assume that $\varphi/M$ settles at $x_0$ at an early stage of the evolution of the universe, leading to an asymptotic behavior with $\alpha = 0$. Up to the additional Goldstone boson $\sigma$ the cosmology is of the standard type. In particular, the value $W(x_0)$ acts as an effective cosmological constant. Realistic cosmologies require $v(x_0)$ to vanish or be very small. From the field equation \cite[eq.~(4.12)]{wetterich1988cosmologies}, $(6 + 4 \omega) \tensor{\sigma}{_;^\mu_\mu} + \frac{1}{2} (\partial W/\partial \sigma) = q^\sigma$, we conclude that $\sigma$ approaches asymptotically a constant value (we assume here $q^\sigma = 0$)
\begin{alignedeqn}
	&\sigma
	= C_1 + C_2 \, \exp(-3 H_0 t)
	&&\text{for } H = H_0,\\
	&\sigma
	= C_1 + C_2 \, t^{1 - 3 \eta}
	&&\text{for } H = \eta t^{-1}.
\end{alignedeqn}
In both cases it can be neglected for late cosmology. The asymptotic value $C_1$ is irrelevant. In conclusion, scale free models can give realistic cosmologies of the standard Friedmann type provided the parameters of the model are such that the cosmological constant $W(x_0)$ vanishes and $x_0$ is very small (gauge hierarchy). The
scale-free version of the Brans-Dicke type action \labelcref{eqn:brans-dicke action} differs from the standard model with fixed Newton's constant only by the presence of the Goldstone boson $\sigma$. For $q^\sigma = 0$ this is irrelevant for late cosmology. Despite its long range, observation of effects due to an exchange or production of this Goldstone boson may be difficult, due to its purely derivative couplings which are suppressed by powers of the Planck mass $M_\text{P}$. We note that for $W(x_0) = 0$ realistic asymptotic cosmology is obtained for arbitrary $\omega$ ($\omega > -\frac{3}{2}$). This may seem puzzling. Inserting the asymptotic value $\varphi = x_0 \chi$, the effective action \labelcref{eqn:brans-dicke action} reduces to the action of the Brans-Dicke theory\cite{brans1961mach} without potential. Standard Brans-Dicke theory, however, is consistent with observation only for $\omega > 500$. The difference comes from the coupling to matter. In the standard Brans-Dicke theory the nucleon and electron masses are treated as intrinsic scales. For varying $\chi$ the observable value of Newton's ``constant'' $G_\text{N} \sim \chi^{-2}$ changes with respect to particle masses. In contrast, a dilatation symmetric quantum theory implies that all particle masses must be proportional to $\chi$ (e.g. $m_e \sim \tilde\varphi \sim \chi$). The observable ratios $m_e^2 \, G_\text{N}$ etc. are therefore static for asymptotic cosmology. In the formalism of \cite[sec.~2]{wetterich1988cosmologies}, the $\chi$-dependence of particle masses leads to a non-vanishing right hand side of the scalar field equation \cite[eq.~(2.7)]{wetterich1988cosmologies} for the matter dominated epoch ($\tilde{q}^\chi \neq 0$). In view of \cite[eq.~(4.14)]{wetterich1988cosmologies} and \cite[eq.~(4.19)]{wetterich1988cosmologies} this is indeed required for the decoupling of the dilaton mode from matter ($q^\sigma = 0$).

It is instructive to understand the Goldstone boson appearing in scale-free models in terms of dilatation invariance. The action \labelcref{eqn:brans-dicke action} with a scale-free potential \labelcref{eqn:effective potential} is invariant under global scale transformations of the fields.
\begin{equation}\label{eqn:global field scale trafo}
	\tilde\varphi
	\to e^\alpha \, \tilde\varphi,
	\qquad
	\chi
	\to e^\alpha \, \chi,
	\qquad
	\tilde{g}_{\mu\nu}
	\to e^{-2 \alpha} \, \tilde{g}_{\mu\nu}.
\end{equation}
In the Weyl-scaled version they read
\begin{equation}
	\varphi
	\to \varphi,
	\qquad
	g_{\mu\nu}
	\to g_{\mu\nu},
	\qquad
	\sigma
	\to \sigma + \alpha \, M.
\end{equation}
The field $\sigma$ is therefore the Goldstone boson which originates from spontaneous breaking of dilatation symmetry for any non-zero value\footnote{The true Goldstone boson contains an admixture of $\tilde\varphi$ to $\chi$ of the order $x_0$.} of $\tilde\varphi$ or $\chi$. It is straightforward to construct the conserved current corresponding to dilatation symmetry. Expressed in the Weyl-scaled fields it reads
\begin{equation}\label{eqn:weyl-scaled dilatation current}
	J_\text{D}^\mu
	= 2 M \, g^{1/2} \biggl\{\biggl(6 + 4 \omega + \frac{\varphi^2}{M^2}\biggr) \tensor{\sigma}{_;^\mu} + \frac{\varphi}{M} \tensor{\varphi}{_;^\mu}\biggr\}.
\end{equation}
Its divergence obviously vanishes as a consequence of the field equations for $\sigma$ derived from \cite[eq.~(4.6)]{wetterich1988cosmologies}. Of course, we could equivalently construct\footnote{For a global infinitesimal transformation $\delta \varphi_i = \alpha \, O_i \, \varphi_i$ with $\alpha$ constant and $O_i$ a differential operator or a constant, and a Lagrange density $\mathcal{L}$ containing terms with up to two derivatives of the fields $\varphi_i$, the symmetry current is
\begin{equation*}
	J^\mu
	= O_i \, \varphi_i \biggl(\frac{\partial \mathcal{L}}{\partial (\partial_\mu \varphi_i)} - \partial_\nu \, \frac{\partial \mathcal{L}}{\partial (\partial_\nu \partial_\mu \varphi_i)}\biggr) + \partial_\nu (O_i \varphi_i) \, \frac{\partial \mathcal{L}}{\partial (\partial_\nu \partial_\mu \varphi_i)} - K^\mu,
\end{equation*}
where $K^\mu$ is obtained from $\delta \mathcal{L} = \alpha \partial_\mu K^\mu$. In our conventions the curvature scalar is
\begin{equation*}
	\tilde{R}
	= \tilde{g}^{\mu\sigma} \, \tilde{g}^{\nu\rho} \bigl(\partial_\mu \partial_\rho \tilde{g}_{\nu\sigma} - \partial_\mu \partial_\sigma \tilde{g}_{\nu\rho}\bigr) + \tensor{\tilde{\Gamma}}{_\mu^\nu_\sigma} \, \tensor{\tilde{\Gamma}}{_\nu^\mu^\sigma} - \tensor{\tilde{\Gamma}}{_\mu^\mu_\sigma} \, \tensor{\tilde{\Gamma}}{_\nu^\nu^\sigma}.
\end{equation*}
The term $\sim 6 \chi \, \partial_\nu \chi$ in \labelcref{eqn:dilatation current} is the contribution from the gravitational part of the action.} the dilatation current from the original action \labelcref{eqn:brans-dicke action}:
\begin{equation}\label{eqn:dilatation current}
	J_\text{D}^\mu
	= 2 \tilde{g}^{1/2} \, \tilde{g}^{\mu\nu} \bigl\{(6 + 4 \omega) \chi \partial_\nu \chi + \tilde\varphi \, \partial_\nu \tilde\varphi\bigr\}.
\end{equation}
For a check of its vanishing divergence we can use the field equations \cite[eqs.~(2.6)~-~(2.8)]{wetterich1988cosmologies} and the identity for a dilatation symmetric potential
\begin{equation}\label{eqn:dilatation symmetric potential identity}
	\chi \, \frac{\partial V}{\partial \chi} + \tilde\varphi \, \frac{\partial V}{\partial \tilde\varphi}
	= 4 V.
\end{equation}
On a manifold\footnote{For manifolds with non-trivial topology Cartesian coordinate systems can be chosen for the different coordinate patches separately.} parametrized by Cartesian coordinates $x^\mu$ we can also formulate a particular general coordinate transformation
\begin{equation}\label{eqn:general coordinate trafo}
	x^\mu
	\to e^{-\alpha} \, x^\mu,
	\qquad
	\tilde{g}_{\mu\nu}
	\to e^{2 \alpha} \, \tilde{g}_{\mu\nu}.
\end{equation}
The infinitesimal transformation is
\begin{alignedeqn}
	x^\mu
	&\to x^\mu + \xi^\mu
	= x^\mu - \alpha \, x^\mu,\\
	\delta \tilde\varphi
	&= -\xi^\lambda \, \partial_\lambda \tilde\varphi
	= \alpha \, x^\lambda \, \partial_\lambda \tilde\varphi,\\
	\delta \tilde{g}_{\mu\nu}
	&= -\partial_\mu \xi^\lambda \tilde{g}_{\lambda\nu} - \partial_\nu \xi^\lambda \tilde{g}_{\mu\lambda} - \xi^\lambda \partial_\lambda \tilde{g}_{\mu\nu}\\
	&= \alpha \, (x^\lambda \, \partial_\lambda + 2) \, \tilde{g}_{\mu\nu}.
\end{alignedeqn}
Combining the transformations \labelcref{eqn:global field scale trafo,eqn:general coordinate trafo} gives another version of dilatation
symmetry where the coordinates instead of the metric are scaled
\begin{alignedeqn}
	&\delta \tilde\varphi
	= \alpha \bigl(x^\lambda \, \partial_\lambda + 1\bigr) \tilde\varphi,
	\qquad
	\delta \tilde{g}_{\mu\nu}
	= \alpha x^\lambda \partial_\lambda \tilde{g}_{\lambda\nu},\\
	&\delta \chi
	= \alpha \bigl(x^\lambda \, \partial_\lambda + 1\bigr) \chi.
\end{alignedeqn}
It is adapted to a special coordinate choice (Cartesian parametrization) and often used to study dilatation transformations on flat space. The corresponding symmetry current in flat space can be constructed from the energy-momentum tensor \cite{coleman1988aspects}.

For dilatation symmetric theories it follows immediately from \labelcref{eqn:dilatation symmetric potential identity} that any extremum of $V(\chi,\tilde\varphi)$ can only occur for vanishing potential $V(\chi_0,\tilde\varphi_0) = 0$. The condition for a vanishing cosmological constant amounts therefore to the requirement that $V$ has a minimum
\begin{equation}
	\frac{\partial V}{\partial \chi}(\chi_0,\tilde\varphi_0)
	= \frac{\partial V}{\partial \tilde\varphi}(\chi_0,\tilde\varphi_0)
	= 0.
\end{equation}

It follows from the general form of the potential \labelcref{eqn:effective potential} that for any extremum at $(\chi_0,\tilde\varphi_0)$ there is also an extremum at $(e^\alpha \, \chi_0,e^\alpha \, \tilde\varphi_0)$. The potential must therefore have a flat direction. Conversely, a flat direction starting from the origin must be at $V = 0$. Flat directions arise when the potential depends only on one particular linear combination of $\tilde\varphi$ and $\chi$. This could be a consequence of some unknown symmetry. We also note that for $V$ convex or bounded from below a zero of $v(\tilde\varphi/\chi)$ is sufficient to produce a flat direction. Any point where $V$ vanishes must be a minimum in this case. We may summarize this section by the following general statement: If the effective potential of a dilatation symmetric quantum theory has a non-trivial minimum, such theories always lead to a Brans-Dicke theory, but with variable particle masses ($m_e \sim \chi$ etc). Such a theory leads to realistic asymptotic cosmology (provided that there is no instability in the kinetic term of the Brans-Dicke scalar, $\omega > -\frac{3}{2}$).

\section[Dilatation anomalies]{Dilatation anomalies\protect\footnote{Parts of this and the next section have been obtained in collaboration with R.D. Peccei and J. Solà and are published in ref.~\cite{peccei1987adjusting}.}}
\label{sec:dilatation anomalies}

Even if we start with a dilatation symmetric action without any mass parameter the properly renormalized quantum field theory sometimes needs the introduction of a mass scale. This occurs if there is no scale invariant way to define the functional measure in the functional integral. In this case renormalization necessarily involves the introduction of a scale, the renormalization scale $\mu$. For scale-dependent (running) renormalized dimensionless couplings $g_i$ the theory must be defined by specifying their values $g_i(\mu)$ at a certain scale $\mu$. In such theories the dilatation symmetry is broken by the quantization -- the theory has a dilatation anomaly. A typical example is pure QCD: The dilatation anomaly is the anomalous trace of the energy-momentum tensor \cite{crewther1972non,chanowitz1972canonical,collins1976renormalization,adler1977energy,collins1977trace,nielsen1977energy}
\begin{equation}
	\tensor{\tilde\vartheta}{_\mu^\mu}
	= \frac{\beta(g_s)}{2 g_s} \, F_a^{\mu\nu} \, F_{\mu\nu}^a,
\end{equation}
with $g_s$ the strong gauge coupling and $\beta(g_s)$ the well-known $\beta$-function of the $SU(3)$ gauge theory.

In the presence of an anomaly the dilatation current is no longer conserved
\begin{equation}
	\partial_\mu J_\text{D}^\mu
	= \Delta
	= \tilde{g}^{1/2} (\tensor{\tilde\vartheta}{_\mu^\mu} + \tilde\vartheta_\text{G}).
\end{equation}
Here we have included a possible dilatation anomaly $\tilde\vartheta_\text{G}$ in the gravitational sector. We can account for the anomaly $\tensor{\tilde\vartheta}{_\mu^\mu}$ in the effective potential for $\tilde\varphi$ and $\chi$. Anomalies introduce an explicit dependence of the effective potential $V$ on the renormalization scale $\mu$,
\begin{equation}\label{eqn:mu-dependent potential}
	V
	= V\bigl(\chi,\tilde\varphi;\mu,g_i(\mu)\bigr).
\end{equation}
For example, in pure QCD the expectation value of $F_a^{\mu\nu} \, F_{\mu\nu}^a$ should be of the order $\Lambda_\text{QCD}^4$. The anomaly gives a constant contribution to $V$, proportional to $\mu^4 \, \exp\bigl(-c/g_s^2(\mu)\bigr)$. In general, the anomaly measures the deviation of the effective potential $V$ from its scale-invariant form
\begin{equation}\label{eqn:potential deviation}
	\tensor{\tilde\vartheta}{_\mu^\mu}
	= 4 V - \chi \, \frac{\partial V}{\partial \chi} - \tilde\varphi \, \frac{\partial V}{\partial \tilde\varphi}.
\end{equation}
Similarly, in the Weyl-scaled version with $\Delta = g^{1/2} (\tensor{\vartheta}{_\mu^\mu} + \vartheta_\text{G})$, one has
\begin{equation}
	\tensor{\vartheta}{_\mu^\mu}
	= \exp\bigl(-\tfrac{4 \sigma}{M}\bigr) \tensor{\tilde\vartheta}{_\mu^\mu}
	= -M \, \frac{\partial W}{\partial \sigma}.
\end{equation}
As a consequence of dilatation anomalies the potential $W$ now depends on $\sigma$ and a has in general non-derivative couplings. For any possible solution with constant and static fields $\tilde\varphi$ and $\chi$ the divergence of the dilatation current \labelcref{eqn:weyl-scaled dilatation current} or \labelcref{eqn:dialtation current appendix} must vanish\footnote{This is similar to the axion \cite{peccei1977cp,peccei1977constraints}. In our case the dilatation anomaly plays an analogous role to the strong $CP$ violating parameter $\bar\vartheta$.} . Such solutions are therefore only possible for values $\varphi_0$, $\sigma_0$ for which the dilatation anomaly $\Delta$ is zero. Let us assume that $\varphi$ has reached a static constant value $\varphi_0$ so that $\partial_\mu \varphi = 0$. The field equation for $\sigma$ is then given (see \cref{sec:appendix}) by
\begin{equation}
	f(\sigma) \, \tensor{\sigma}{_;^\mu_\mu} + g(\sigma) \, \tensor{\sigma}{_;^\mu} \, \tensor{\sigma}{_;_\mu}
	= \frac{1}{M} \, \vartheta(\sigma),
\end{equation}
where $\vartheta(\sigma)$ is the anomaly $\tensor{\vartheta}{_\mu^\mu} + \vartheta_\text{G}$ with all terms containing derivatives of $\sigma$ subtracted. Obviously, the anomaly $\vartheta(\sigma)$ acts as a driving force for $\sigma$.

Solutions with constant static $\sigma = \sigma_0$ require
\begin{equation}\label{eqn:static sigma requirement}
	\vartheta(\sigma_0)
	= 0.
\end{equation}
These solutions are stable only if the mass term for the excitation is positive (or vanishes)
\begin{equation}\label{eqn:stability condition}
	m_\sigma^2
	= -\frac{1}{M \, f(\sigma_0)} \, \frac{\partial \vartheta}{\partial \sigma} (\sigma_0)
	\geq 0.
\end{equation}
The effective cosmological constant for $\sigma = \sigma_0$ is $W(\sigma_0)$. It should vanish for any realistic cosmology and we must require
\begin{equation}\label{eqn:vanishing effective cosmological constant}
	W(\sigma_0)
	= 0.
\end{equation}
Otherwise the universe approaches asymptotically an exponential expansion ($W(\sigma_0) > 0$) or a catastrophic contraction ($W(\sigma_0) < 0$). Such a behavior would be much more singular than the Brans-Dicke cosmologies with $V_0 \neq 0$ discussed in \cite[sec.~3]{wetterich1988cosmologies},
which approach flat space asymptotically.
For solutions fulfilling \labelcref{eqn:vanishing effective cosmological constant} the actual value of $\sigma_0$ is irrelevant. In addition, we
have a freedom in the definition of $\sigma$ and $M$ since $\chi = M \, \exp(\sigma/M)$ remains unchanged under the transformation
\begin{equation}
	\sigma^\prime
	= e^{-\hat\alpha} (\sigma + \hat\alpha \, M),
	\qquad
	M^\prime
	= e^{-\hat\alpha} \, M.
\end{equation}
We use this freedom to set $\sigma_0 = 0$. It is the corresponding value of $M$ which is related to Newton's constant \cite[eq.~(4.2)]{wetterich1988cosmologies}.

Let us look at the role of a possible dilatation anomaly in the gravitational sector (see \cref{sec:appendix}) in view of the combined conditions \labelcref{eqn:static sigma requirement,eqn:vanishing effective cosmological constant}: From the definition of $\vartheta(\sigma)$ (see \cref{eqn:f+g}) one sees immediately that they require
\begin{equation}
	\tensor{\vartheta}{_\mu^\mu}(\sigma_0)
	= 0.
\end{equation}
We conclude that the dilatation anomaly in the gravitational sector does not play an essential role for such theories. For the remainder of this paper we will use the simplification $\vartheta_\text{G} = 0$ ($h = 1$).

We can express \labelcref{eqn:static sigma requirement,eqn:vanishing effective cosmological constant} as conditions on $V$:
\begin{align}\label{eqn:fine tuning condition}
	&4 V(\chi_0,\tilde\varphi_0) - \chi \, \frac{\partial V}{\partial \chi}(\chi_0,\tilde\varphi_0) - \tilde\varphi \, \frac{\partial V}{\partial \tilde\varphi}(\chi_0,\tilde\varphi_0)
	= 0,\notag\\
	&V(\chi_0,\tilde\varphi_0)
	= 0.
\end{align}
The minimum of $V$ must be at zero. This corresponds to the usual fine tuning condition for the cosmological constant. In our context, however, this has a perhaps more physical interpretation: We require a model with the property that the trace of the energy-momentum tensor in the vacuum with static $\chi_0$ and $\tilde\varphi_0$ ($\tensor{\tilde{T}}{_\mu^\mu} = 4 V$) is given by its anomalous part $\tensor{\tilde\vartheta}{_\mu^\mu}$,
\begin{equation}\label{eqn:cosmon condition}
	\tensor{\tilde{T}}{_\mu^\mu}(\chi_0,\tilde\varphi_0)
	= \tensor{\tilde\vartheta}{_\mu^\mu}(\chi_0,\tilde\varphi_0).
\end{equation}
If a static $\chi_0$ exists $\tensor{\tilde\vartheta}{_\mu^\mu}(\chi_0)$ must vanish and if in addition the stability condition \labelcref{eqn:stability condition} holds, the cosmology necessarily approaches flat space asymptotically. The field $\sigma$ is then called a ``cosmon'' \cite{peccei1987adjusting}. Its dynamics drives the anomalous trace $\tensor{\tilde\vartheta}{_\mu^\mu}$ and by \labelcref{eqn:cosmon condition} the cosmological constant to zero. Realistic cosmology of the standard type is obtained provided the energy stored in coherent oscillations of the cosmon never exceeds the radiation energy during the usual radiation dominated epoch \cite{preskill1983cosmology,abbott1983cosmological,dine1983not}. This depends on ``initial conditions'' for the amplitude of cosmon oscillations. Since the cosmon always couples to the anomalous trace $\tensor{\tilde\vartheta}{_\mu^\mu}$ its evolution in the history of the universe may be rather complicated, especially during phase transitions when condensates form. This issue certainly merits further study.

Since $\vartheta$ depends only on $\sigma/M$ we can immediately conclude that the cosmon mass \labelcref{eqn:stability condition} is of the order
\begin{equation}
	m_\sigma^2
	\approx \frac{m^4}{M^2}.
\end{equation}
Here $m$ is typically the largest characteristic scale produced by the anomaly. In our approach, $m$ should be at most of the order of the Fermi scale $\varphi_0 \approx \SI{174}{\giga\electronvolt}$. A lower bound would be given by $\Lambda_\text{QCD}$ if strong interactions were a fundamental theory valid to arbitrarily short distances:
\begin{equation}\label{eqn:cosmon mass lower bound}
	m_\sigma
	\gtrsim \frac{\Lambda_\text{QCD}^2}{M_\text{P}}
	\approx \SI{2e-11}{\electronvolt}.
\end{equation}
This would give an upper bound on the range of about \SI{10}{\kilo\meter}. Possible detection of an intermediate range cosmon force is discussed in ref.~\cite{peccei1987adjusting}. We will see below that $m$ depends crucially on the short distance behavior of the full (unified) theory. A cosmon mass quite different from \labelcref{eqn:cosmon mass lower bound} should therefore not be excluded at this point.

\section{The cosmon condition}
\label{sec:cosmon condition}

Let us concentrate on theories where all intrinsic mass scales (such as $m$) appear only through the running of dimensionless couplings. Physical quantities only depend on renormalized dimensionless coupling constants $g_i(\mu)$ defined at some renormalization scale $\mu$. Except for $\mu$ no explicit mass parameter should appear in the theory. In such theories the ``cosmon condition'' \labelcref{eqn:cosmon condition} requires a non-trivial connection between the long distance and short distance properties of a theory. Let us assume for simplicity that the fundamental theory has only one dimensionless running coupling constant $g(\mu)$ and therefore no free adjustable dimensionless parameter. The effective potential \labelcref{eqn:mu-dependent potential} must have the form
\begin{equation}
	V
	= \tilde\varphi^4 \, v\bigl(\chi/\mu,\tilde\varphi/\mu;g(\mu)\bigr).
\end{equation}
and one finds for the dilatation anomaly
\begin{equation}\label{eqn:dilatation anomaly}
	\tensor{\tilde\vartheta}{_\mu^\mu}
	= \mu \, \frac{\partial V}{\partial \mu}.
\end{equation}
Using the independence of $V$ on the choice of the renormalization scale $\mu$ (the renormalization group equation) one also obtains
\begin{equation}
	\tensor{\tilde\vartheta}{_\mu^\mu}
	= -\beta \, \frac{\partial V}{\partial g(\mu)},
	\qquad
	\beta
	= \mu \, \frac{\partial g(\mu)}{\partial \mu}.
\end{equation}
In particular, there is no anomaly if the fundamental coupling $g$ is not running ($\beta = 0$), despite the fact that all low-energy couplings may be scale-dependent. The cosmon condition reads
\begin{equation}
	\mu \, \frac{\partial V}{\partial \mu}(\chi_0,\tilde\varphi_0)
	= 4 V(\chi_0,\tilde\varphi_0)
	= 0.
\end{equation}

On the other hand the cosmon condition is related to the $\chi$-dependence of the effective potential
\begin{equation}\label{eqn:eff potential chi-dependence}
	\tilde\vartheta_\chi(\chi_0,\tilde\varphi_0)
	= -\chi \, \frac{\partial V}{\partial \chi}(\chi_0,\tilde\varphi_0)
	= 0.
\end{equation}
In flat space this is just the condition that static fields $\chi_0$, $\tilde\varphi_0$ must correspond to an extremum of the potential. In the presence of curvature, however, $\chi_0$ is determined (for $(\partial V/\partial \tilde\varphi)(\chi_0,\tilde\varphi_0) = 0$) by $\chi_0 \, \frac{\partial V}{\partial \chi}(\chi_0,\tilde\varphi_0) = 4 V(\chi_0,\tilde\varphi_0)$ and \labelcref{eqn:eff potential chi-dependence} is a non-trivial condition on the theory\footnote{We note that $\tilde\vartheta_\chi(\chi,\tilde\varphi_0) = 0$ for all $\chi$ would lead to the Brans-Dicke theories discussed in \cite[sec.~3]{wetterich1988cosmologies}, which are only realistic for $V(\chi,\tilde\varphi_0) = 0$.}. For flat space and static fields one has the general identity $\langle\tensor{\tilde\vartheta}{_\mu^\mu}\rangle = \langle\tensor{T}{_\mu^\mu}\rangle$. This explains the historical origin of the name ``anomalous trace of the energy-momentum tensor'' for the dilatation anomaly. Indeed, once all VEVs are expressed in terms of $\mu$ one necessarily has $V_0 = c_0 \, \mu^4$. Using $\partial V/\partial \varphi_i = 0$ for all fields $\varphi_i$ one obtains
\begin{alignedeqn}
	\tensor{\tilde{T}}{_\mu^\mu}
	&= 4 V_0
	= \mu \, \frac{\dif V}{\dif \mu}
	= \mu \, \frac{\partial V}{\partial \mu} + \mu \, \frac{\partial \varphi_i}{\partial \mu} \, \frac{\partial V}{\partial \varphi_i}\\
	&= \mu \, \frac{\partial V}{\partial \mu}
	= \tensor{\tilde\vartheta}{_\mu^\mu}.
\end{alignedeqn}
Inclusion of gravity only permits to conclude an identity for the partial $\mu$-derivative of $\tensor{T}{_\mu^\mu}$ (for fixed VEVs of $\varphi_i$),
\begin{equation}
	\langle \tensor{\tilde\vartheta}{_\mu^\mu}\rangle
	= \frac{1}{4} \mu \, \frac{\partial}{\partial \mu} \langle\tensor{T}{_\mu^\mu}\rangle.
\end{equation}

As discussed in the introduction of \cite{wetterich1988cosmologies}, there are several types of possible sources for the $\chi$-dependence of the effective potential. It may arise from mass-type terms in the effective potential for $\tilde\varphi$, like
\begin{equation}\label{eqn:eff potential mass term}
	\Delta V_\chi
	= \varepsilon \, \chi^2 \tilde\varphi^2 + \kappa \, \chi^4.
\end{equation}
Such terms give no contribution to $\tensor{\tilde\vartheta}{_\mu^\mu}$. Following the ideas of \cite{wetterich1988cosmologies} these terms should be absent (or their coefficients be very small). A second source, more related to the spirit of \cite{wetterich1988cosmologies}, comes from the fact that the standard $SU(3) \times SU(2) \times U(1)$ model is not expected to be valid up to infinitely high energies. At short distances we expect that the theory shows a higher symmetry, possibly connected to grand unification, higher dimensions or strings. This symmetry must be spontaneously broken and it is natural to associate the corresponding symmetry breaking scale $M_x$ with the expectation value of $\chi$ and therefore with $M_\text{P}$:
\begin{equation}
	M_x
	= \hat\gamma \, \chi.
\end{equation}
The ratio $\hat\gamma$ should not be very far from unity and we will take $\hat\gamma = 1$ unless stated otherwise. The change of $\beta$-functions at $M_x$ produces a $\chi$-dependence of the effective potential. As an illustration we give the $\chi$-dependence of $\Lambda_\text{QCD}$ in an $SU(5)$ theory in the one-loop approximation:
\begin{equation}
	\Lambda_\text{QCD}
	= \exp\bigl[\bigl(2 b_3 \, g_0^2\bigr)^{-1}\bigr] \, \mu^{b_5/b_3} \, (\hat\gamma \chi)^{1 - b_5/b_3}.
\end{equation}
Here the $SU(5)$ coupling is defined at a renormalization scale $\mu \gg \chi$, $g_0 = g(\mu)$, and $b_5$, $b_3$ are the usual coefficients of the $g^3$ term in the $\beta$-function ($\beta = b \, g^3$) above and below the scale $M_x$. (We have neglected fermion-mass thresholds.)

More generally, we can understand this contribution to the $\chi$-dependence of $V$ in terms of the renormalization group equations. The scale-invariant version of the standard model admits Higgs mass terms in the action only in the form $\Delta V_\chi$ \labelcref{eqn:eff potential mass term}. It has therefore only dimensionless couplings. Let us denote their values at the symmetry breaking scale $M_x$ by $g_j(\chi)$. Below $M_x$, the scale dependence of $g_j$ is given by the usual renormalization group equations of the standard model with $\beta$-functions $\beta_j$. Above $M_x$ the evolution equations change and are determined by different $\beta$-functions $\hat\beta_j$. (In a fundamental theory the $\hat\beta_j$ are all related to $\beta$ in \cref{eqn:dilatation anomaly}.) Formally one has
\begin{alignedeqn}
	&\beta_j(\chi)
	= \chi \, \frac{\partial g_j(\chi)}{\partial \chi}\biggr|_{\tilde\varphi,\, g_j(\tilde\varphi) \text{ fixed}}\\
	&\hat\beta_j(\chi)
	= \chi \, \frac{\partial g_j(\chi)}{\partial \chi}\biggr|_{\mu,\, g_j(\mu) \text{ fixed}}
	\quad (\tilde\varphi \ll \chi \ll \mu).
\end{alignedeqn}
We can express the effective potential entirely in terms of the $g_j(\chi)$:
\begin{equation}
	V
	= \tilde\varphi^4 \, \bar{v}\bigl(\tilde\varphi/\chi;g_j(\chi)\bigr).
\end{equation}
This implies
\begin{alignedeqn}
	&\tensor{\tilde\vartheta}{_\mu^\mu}
	= \hat\beta_j \, \frac{\partial \bar{v}}{\partial g_j(\chi)} \, \tilde\varphi^4,\\
	&\tilde\vartheta_\chi
	= -\biggl(\chi \, \frac{\partial \bar{v}}{\partial \chi} + \hat\beta_j \, \frac{\partial \bar{v}}{\partial g_j(\chi)}\biggr)\tilde\varphi^4.
\end{alignedeqn}
Neglecting the explicit $\chi$-dependent contributions $\Delta V_\chi$ \labelcref{eqn:eff potential mass term} we can determine the $\chi$-dependence of $\bar{v}$ by the ``low energy'' renormalization group equations
\begin{equation}
	\chi \, \frac{\partial \bar{v}}{\partial \chi} + \beta_j \, \frac{\partial \bar{v}}{\partial g_j(\chi)}
	= 0.
\end{equation}
In this formulation, the cosmon condition is equivalent to a non-trivial matching condition for the low energy and high energy $\beta$-functions:
\begin{equation}\label{eqn:matching condition}
	\beta_j \, \frac{\partial \bar{v}}{\partial g_j(\chi)}\biggr|_{\tilde\varphi_0,\chi_0}
	= \hat\beta_j \, \frac{\partial \bar{v}}{\partial g_j(\chi)}\biggr|_{\tilde\varphi_0,\chi_0}
	= 0.
\end{equation}
Such a condition is certainly quite suggestive, but not well understood. In a theory with free adjustable parameters it seems not particularly difficult to choose parameters such that \labelcref{eqn:matching condition} holds for some $\chi_0$. In a fundamental theory without adjustable parameters, however, the condition \labelcref{eqn:matching condition} (or its generalization for $\Delta V_\chi \neq 0$) would be a remarkable property.

\section{An anomalous renormalization group equation for the cosmological ``constant''}
\label{sec:anomalous renormalization group equation}

The simplest solution of having both $\tensor{\tilde\vartheta}{_\mu^\mu}$ and $V$ simultaneously vanishing for some value $\chi_0$ would be that they are proportional to each other:
\begin{equation}
	\mu \, \frac{\partial V}{\partial \mu}
	= A \, V.
\end{equation}
In this case the renormalization group equation for $V$ would be entirely determined by the anomalous dimension $A$. The dimensionless quantity $A$ may depend on the dimensionless coupling constants of the theory. This is what happens in a pure $\varphi^4$-theory\footnote{The one-loop approximation to the scalar potential in the dilatation symmetric standard model has been discussed recently by Buchmüller and Dragon \cite{buchmuller1989dilatons}. Their method implicitly assumes an extension of the standard model to infinitely short distances. The results coincide with our formalism for $\tilde{g}_{\mu\nu} = \eta_{\mu\nu}$. For $\tilde{g}_{\mu\nu} = c \, \eta_{\mu\nu}$, however, Buchmüller and Dragon use a regularization and renormalization which differs from ours. It gives different results for different coordinate parametrizations of Minkowski space and leads to a term $\tilde{g}^{1/2} \, \ln \tilde{g}$, whose meaning and consistency is not immediately apparent.},
\begin{alignedeqn}\label{eqn:rg egn A-determined}
	V
	&= \frac{1}{2} \, \lambda \biggl(\frac{\tilde\varphi}{\mu},\lambda(\mu)\biggr) \tilde\varphi^4,\\
	\mu \, \frac{\partial V}{\partial \mu}
	&= \frac{1}{2} \, \mu \frac{\partial \lambda}{\partial \mu} \, \tilde\varphi^4
	= -\frac{1}{2} \, \tilde\varphi \frac{\partial \lambda}{\partial \tilde\varphi} \, \tilde\varphi^4\\
	&= -\frac{1}{2} \beta_\lambda \tilde\varphi^4
	= -\frac{\beta_\lambda}{\lambda} \, V.
\end{alignedeqn}
In analogy, suppose for a moment that one could write the effective potential of a fundamental theory with only one running coupling constant $g$ as
\begin{equation}\label{eqn:single running coupling}
	V
	= g^2(\chi) \, \hat{V}
	= g^2\bigl(\chi/\mu,g(\mu)\bigr) \, \hat{V},
\end{equation}
with $\hat{V}$ independent of $\mu$. The only fundamental mass scale of such a theory is the scale of anomalous dilatation symmetry breaking generated by the running of $g$. Therefore the quantity $V$ should be a quartic polynomial in the various (perhaps infinitely many) scalar fields of this theory\footnote{There is an appropriate generalization for fermions or other bosonic fields. Condensates of such fields can again be expressed in terms of scalar operators.}. In this case dilatation symmetry breaking would be entirely described by the anomalous dimension of $V$:
\begin{alignedeqn}
	\mu \, \frac{\partial V}{\partial \mu}
	&= \mu \, \frac{\partial g^2}{\partial \mu} \, \hat{V}
	= -2 g \, \chi \, \frac{\partial g}{\partial \chi} \, \hat{V}\\
	&= -2 g \, \hat\beta \hat{V}
	= -\frac{2 \hat\beta}{g} \, V
	= A \, V.
\end{alignedeqn}
On the other hand, scale ratios like $\tilde\varphi/\chi$ would only depend on the properties of $\hat{V}$ and be independent of the scale of dilatation symmetry breaking.

Of course, the assumption \labelcref{eqn:single running coupling} is very strong and we do not expect it to hold except for very simple theories. Nevertheless, the property that scale ratios like $\tilde\varphi/\chi$ are independent of dilatation anomalies should always be a very good approximation if the scale $m$ characteristic for anomalous dilatation symmetry breaking is much smaller than $\tilde\varphi$ and $\chi$. Consider now a relative minimum of $V$ with respect to all fields except $\chi$ ($\partial V/\partial \tilde\varphi = 0$) and denote the corresponding value of the effective potential by $V_0(\chi)$. Instead of \labelcref{eqn:single running coupling} we will only assume
\begin{equation}\label{eqn:cosmological constant rg equation}
	\mu \, \frac{\partial V_0}{\partial \mu}
	= A \, V_0.
\end{equation}
Such a behavior would be suggested if $V_0(\chi)$ is the only relevant quantity with dimension of mass. We may call \labelcref{eqn:cosmological constant rg equation} the renormalization group equation for the cosmological ``constant''. More generally, if $A$ is a function of $\chi$, it only can depend on the ratio $m/\chi$ where $m$ is the physical scale generated by the dilatation anomaly. (Remember that $\tensor{\tilde\vartheta}{_\mu^\mu}$ and $V_0$, and therefore $A$, are physical quantities which must be independent under a simultaneous change of $\mu$ and $g(\mu)$.) It will be sufficient for our purpose if $A(m/\chi)$ approaches a constant $A \neq 0$ in the limit where $m/\chi$ goes to zero.

A renormalization group equation of the type \labelcref{eqn:cosmological constant rg equation} has important consequences. Consider the case where $A$ can be approximated by a constant. For an asymptotically free theory $A$ should be positive. Using \labelcref{eqn:potential deviation,eqn:dilatation anomaly} one obtains
\begin{equation}\label{eqn:natural potential}
	V_0
	= a_0 \biggl(\frac{m}{\chi}\biggr)^A \, \chi^4.
\end{equation}
For $A < 4$ the anomaly $\tensor{\tilde\vartheta}{_\mu^\mu} = A \, V_0$ vanishes for $\chi_0 = 0$. In this case all scales disappear for the static solution $\chi_0 = \tilde\varphi_0 = 0$ and dilatation symmetry becomes restored for such a solution. For $A > 4$ the trace anomaly only vanishes for $\chi \to \infty$. There is no finite static $\chi_0$ fulfilling \labelcref{eqn:static sigma requirement} and we expect that $\chi$ moves asymptotically to infinity. We will discuss the corresponding cosmology in the next section. The intermediate case $A = 4$ corresponds to the discussion in \cite[sec.~3]{wetterich1988cosmologies} with the important difference that now $\tilde\varphi/\chi$ instead of $\tilde\varphi$ is kept constant. For all cases the renormalization group equation is inconsistent with \labelcref{eqn:static sigma requirement} for finite non-vanishing $\chi_0$ unless $a_0 = 0$.

Nevertheless, for $A > 4$, the cosmon condition \labelcref{eqn:fine tuning condition} is fulfilled asymptotically for $\chi \to \infty$. For large enough $\chi$ the cosmological constant $V_0$ becomes arbitrarily small. Any non-vanishing positive $a_0$ may be absorbed by a redefinition of $m$ and we take $a_0 = 1$. The scale $m$ may then be identified with the characteristic scale generated by anomalous dilatation symmetry breaking. In a fundamental theory it is the only intrinsic scale and sets the units for all other operators with dimension of mass. In units where today's value of $\chi$ is $\chi_0 = \SI{1.7e18}{\giga\electronvolt}$ the scale $m$ is bounded by today's observed value for the Hubble parameter,
\begin{alignedeqn}
	&H_0
	= 2 h_0 \cdot \SI{e-33}{\electronvolt},\\
	&V_0
	\leq \frac{3 H_0^2 M_\text{P}^2}{8 \pi}
	= (\SI{3e-3}{\electronvolt})^4 \, h_0^2,\\
	&m
	= \cramped{\biggl(\frac{V_0}{\chi_0^4}\biggr)^\frac{1}{A}} \chi_0
	\leq \num{1.7}^{1 + \frac{4}{A}} \, h_0^\frac{2}{A} \cdot 10^{27 - \frac{120}{A}} \si{\electronvolt}.
\end{alignedeqn}
For $A > 4$ the bound on $m$ is bigger than $3 h_0^{1/2} \cdot \SI{e-3}{\electronvolt}$ and it approaches this value for $A \to 4$.

One may ask if it is reasonable that today's value of $V_0$ is in the range $\sim \SI{e-46}{\giga\electronvolt^4}$ or smaller although individual contributions from QCD and weak symmetry breaking could have a characteristic size of \SIrange{e-2}{e8}{\giga\electronvolt^4}. Let us first discuss this question for the contributions to the dilatation anomaly $\tensor{\tilde\vartheta}{_\mu^\mu}$. First of all we note that ``individual contributions'' from different sectors of the theory are not really well defined. Weak interactions and QCD are not independent. (For example, quark masses arise from weak symmetry breaking and play a role in QCD.) As a consequence $\tensor{\tilde\vartheta}{_\mu^\mu}$ (and $V_0$) is not simply an addition of a pure weak and a pure QCD piece. If we nevertheless decide on some definition for the individual contributions to $\tensor{\tilde\vartheta}{_\mu^\mu}$, they will typically reflect how the effective potential changes if one varies a certain degree of freedom while keeping the others fixed. This is connected to the physics determining scale ratios. Scale ratios like $\Lambda_\text{QCD}/\tilde\varphi$ typically depend on dimensionless couplings and may be unrelated to the fate of dilatation symmetry. Individual contributions to $\tensor{\tilde\vartheta}{_\mu^\mu}$ may therefore be large (for example $\sim \tilde\varphi^4$) even for a theory where dilatation symmetry has no anomaly at all. The total dilatation anomaly ~ is related to different physics, namely the connection between the overall scale of VEVs and the intrinsic scale $m$. It can be much smaller than the individual contributions. These must simply cancel if the theory either has no anomaly, or if the anomaly vanishes in the vacuum as a result of the dynamics of the $\chi$-field ($\tensor{\tilde\vartheta}{_\mu^\mu}(\chi_0) = 0$), or if $\tensor{\tilde\vartheta}{_\mu^\mu}$ vanishes asymptotically for $m/\chi \to 0$. In the latter case the existence of two different scales $m$ and $\chi$ (characterizing intrinsic and spontaneous breaking of dilatation symmetry) is crucial.

The minimum value $V_0$ of the effective potential is a quantity connected with a scale ratio, namely $\tilde{R}/\chi^2$. A priori it is therefore not necessarily related to the fate of the overall scale and could be of the order of its individual contributions even if $\tensor{\tilde\vartheta}{_\mu^\mu}$ is much smaller. For theories which establish a connection between $V_0$ and $\tensor{\tilde\vartheta}{_\mu^\mu}$, however, the situation is different. For $\tensor{\tilde\vartheta}{_\mu^\mu} = A \, V_0$ \labelcref{eqn:cosmological constant rg equation} the physics responsible for a small $\tensor{\tilde\vartheta}{_\mu^\mu}$ also leads to a small value $V_0$, independent of the size of its individual contributions. A potential $V_0$ of the form \labelcref{eqn:natural potential} would then be natural even if individual contributions to $V_0$ are of the order $\tilde\varphi^4$ or even larger. We note in particular that no small dimensionless coupling appears in \labelcref{eqn:natural potential}. The smallness of today's value of $V_0$ directly follows from the small ratio $m/\chi$. We still have to ask in this case if a value of $\chi$ much larger than $m$ is natural. Already the most naive consideration for a theory with only one mass scale $m$ (and without very small dimensionless quantities) would suggest that the possible values for an asymptotic solution for $\chi$ should correspond to $\chi \approx m$, $\chi = 0$ or $\chi \to \infty$. We will see in the next section that it is the latter case which is realized for $V_0$ of the type \labelcref{eqn:natural potential}. The smallness of today's value of $m/\chi$ then follows naturally as a result of the age of the universe.

A fundamental theory leading to the evolution equation \labelcref{eqn:cosmological constant rg equation} can therefore
predict a very small value for both $\tensor{\tilde\vartheta}{_\mu^\mu}$ and $V_0$ (as observed today) as a dynamical result of the evolution of the universe. The cosmon conditions \labelcref{eqn:static sigma requirement,eqn:stability condition,eqn:fine tuning condition,eqn:eff potential chi-dependence,eqn:matching condition} must be fulfilled today within a very good approximation! This places a restriction on the allowed values of the dimensionless couplings of the effective low energy theory. This condition, $V_0 < H_0^2 \, M_\text{P}^2$, is equivalent to the usual fine tuning condition for the cosmological constant. In our case, however, it does not follow as a result of a special choice of fundamental coupling constants but rather as a consequence of the short distance behavior of the theory leading to \labelcref{eqn:cosmological constant rg equation}. For any ground state consistent with \labelcref{eqn:cosmological constant rg equation} the dimensionless couplings must adjust to give a tiny value $V_0$ today. (For the example of a higher dimensional theory the shape of internal space must adjust correspondingly.) If the most general terms in $\Delta V_\chi$ \labelcref{eqn:eff potential mass term} would be present this only would restrict the allowed value of the unobservable coupling $\kappa$. If we discard $\Delta V_\chi$ according to the spirit described in the introduction of \cite{wetterich1988cosmologies}, the model becomes much more predictive. For a short distance behavior \labelcref{eqn:cosmological constant rg equation} and $\Delta V_\chi = 0$ the perturbative approximation for the effective potential of the Higgs doublet leads to a prediction for both the Higgs boson mass and the top quark mass (in case of three generations). Indeed, since for the vacuum $\partial V/\partial \tilde\varphi = 0$ and $\partial V/\partial \chi \approx 0$ holds, the dilatation anomaly for the weak Higgs doublet is given by $V_0 \approx -\frac{1}{2} \, \beta_\lambda \, \tilde\varphi^4$ \labelcref{eqn:rg egn A-determined}. This should be at most of the order of the QCD contribution $\sim \Lambda_\text{QCD}$. The $\beta$-function for the quartic scalar coupling must therefore be very small. The positive contributions to $\beta_\lambda$ arising from the gauge interactions must cancel the negative contributions from the Yukawa coupling of the top quark. If the one-loop approximation for the effective scalar potential \cite{coleman1973radiative} is valid one obtains a top quark mass of \SI{80}{\giga\electronvolt}. For this value of $m_t$ the mass of the physical Higgs boson is unusually small, below \SI{1}{\giga\electronvolt}.

\section{Cosmology with time variation of the cosmological ``constant''}
\label{sec:cosmology with time variation}

In this section we study the cosmology with a potential $V_0(\chi) = (m/\chi)^A \chi^4$. The discussion is analogous to the model of \cite[sec.~3]{wetterich1988cosmologies} (which is recovered for $A = 4$)\footnote{This holds for $n = 4$. For $n = 3$ the approximation $q^\sigma = 0$ is not equivalent to $\tilde{q}^\chi = 0$ in \cite[sec.~3]{wetterich1988cosmologies}.}. However, here we assume that $\tilde\varphi/\chi$ instead of $\chi$ is time-independent. We will use the Weyl-scaled field equations \cite[eq.~(4.18)]{wetterich1988cosmologies} with $\dot\varphi = 0$. The potential reads
\begin{equation}\label{eqn:const varphi potential}
	W_0(\sigma)
	= \biggl(\frac{m}{M}\biggr)^A M^4 \, \exp\biggl(-\frac{A \sigma}{M}\biggr).
\end{equation}
We neglect for a moment incoherent fluctuations ($\rho = p = q^\sigma = q^\varphi = 0$). The field equation for $\sigma$ is
\begin{alignedeqn}
	&\frac{\ddot\sigma}{M} + 3 H \, \frac{\dot\sigma}{M}
	= c \, \exp(-A \sigma/M),\\
	&c
	= \frac{A}{4(3 + 2 \omega)} \, \biggl(\frac{m}{M}\biggr)^A M^2.
\end{alignedeqn}
For $H(t)= \eta \, t^{-1}$ this has a particular solution
\begin{equation}\label{eqn:cosmoglogical evolution}
	\sigma(t)
	= \sigma(t_0) + \frac{2 M}{A} \, \ln \frac{t}{t_0},
\end{equation}
provided
\begin{equation}
	3 \eta - 1
	= \frac{1}{2} A \, c \, t_0^2 \, \exp\bigl(-A \frac{\sigma(t_0)}{M}\bigr).
\end{equation}
The remaining field equations \cite[eq.~(4.18)]{wetterich1988cosmologies} are fulfilled for
\begin{equation}\label{eqn:y def}
	\eta
	= \frac{4 (3 + 2 \omega)}{A^2}
	\equiv Y.
\end{equation}
Let us now include relativistic ($n = 4$) or non-relativistic ($n = 3$) matter (still keeping $q^\sigma = 0$). It is easy to see that for $Y > 2/n$ all matter effects become asymptotically negligible since $\rho$ decreases $\sim t^{-n Y}$. One would end with a universe containing essentially only coherent motions of the $\sigma$ field coupled to gravity. Asymptotic cosmology is given by \labelcref{eqn:cosmoglogical evolution,eqn:y def}. For $Y < 2/n$, however, the asymptotic solutions look different. One now finds
\begin{align}
	&\eta
	= 2/n,
	\qquad
	\rho
	= \rho_0 \, t_0^2 \, t^{-2},\\
	&\frac{\rho_0 \, t_0^2}{6 M^2}
	= \frac{4}{n^2} \bigl(1 - \tfrac{1}{2} n Y\bigr).
	\label{eqn:asymptotic solution}
\end{align}
Depending on $Y$ we therefore have the following possibilities for the asymptotic behaviour: for $Y > \frac{2}{3}$ the universe is $\sigma$-dominated ($\rho$ can be neglected). For $\frac{1}{2} < Y < \frac{2}{3}$ the universe is $\sigma$-dominated during the period when matter is dominantly relativistic. When matter becomes non-relativistic the universe turns to the usual behavior $a \sim t^{2/3}$. Finally, for $Y < \frac{1}{2}$ both the radiation dominated and the matter dominated period have the standard expansion laws $a \sim t^{1/2}$ and $a \sim t^{2/3}$, respectively.

It is instructive to interpret these asymptotic solutions in terms of a cosmological ``constant'' $\lambda$ which varies with time. There are two contributions of the $\sigma$ field to the energy-momentum tensor: One comes from the potential $W(\sigma)$ and the other from the kinetic term $\sim \dot\sigma^2$. The definition of the cosmological constant, energy density and pressure is ambiguous. One possibility would be to identify $\lambda = W$, $\rho_\sigma = p_\sigma = (6 + 4 \omega) \dot\sigma^2$. This has the disadvantage that in presence of both the $\sigma$ field and matter (radiation) the ratio between $p$ and $\rho$ would be different for the two components of the energy-momentum tensor. We therefore adopt the definition \cite{reuter1987time}
\begin{equation}
	p_\sigma
	= \bigl(\tfrac{1}{3} \, n - 1\bigr) \rho_\sigma,
\end{equation}
with $n = 4$ or $3$ for the radiation-dominated or matter-dominated period, respectively. This determines which part of the kinetic term is counted in the cosmological constant
\begin{alignedeqn}
	&\lambda
	= W - \bigl(\tfrac{12}{n} - 2\bigr) (3 + 2 \omega) \dot\sigma^2,\\
	&\rho_\sigma
	= \frac{12}{n} (3 + 2 \omega) \dot\sigma^2.
\end{alignedeqn}
The energy-momentum tensor is
\begin{equation}
	T_{00}
	= \lambda + \rho_\sigma + \rho,
	\qquad
	T_{ij}
	= (\lambda - p_\sigma - p) g_{ij}.
\end{equation}
We can express the time derivative of $\lambda$ in terms of $\lambda$, $H$ and $\dot\sigma$,
\begin{alignedeqn}
	\dot\lambda
	&= -6 \lambda \, \frac{A \dot\sigma}{n M}\\
	&\hphantom{{}=} + 12 \bigl(\tfrac{6}{n} - 1\bigr) (3 + 2 \omega) \dot\sigma^2 \biggl(H -  \frac{A \dot\sigma}{nM}\biggr).
\end{alignedeqn}
This cosmology is of the general type discussed in ref.~\cite{reuter1987time}\footnote{For other attempts to obtain a vanishing cosmological constant as a result of dynamics see refs.~\cite{dolgov1983very,wilczek1983erice,antoniadis1984cosmological,abbott1985mechanism,zee1985remarks,nilles1985florence,barr1986survey}.}. For the asymptotic behavior one has $A \dot\sigma/n M = H$ if $Y < 2/n$. Then $\lambda$ decreases faster than $t^{-2}$ and becomes negligible. In the language of ref.~\cite{reuter1987time} we have a $\rho$-dominated universe but the energy density contains now an additional contribution $\rho_\sigma$ compared to standard cosmology. The energy of coherent $\sigma$-motion $\rho_\sigma$ approaches $ 3 W$ ($2 W$) for $n = 4$ ($3$). Its relative contribution to the energy density is (see \labelcref{eqn:asymptotic solution}).
\begin{equation}\label{eqn:sigma-motion contribution}
	\frac{\rho_\sigma}{\rho + \rho_\sigma}
	= \frac{1}{2} \, n \, Y.
\end{equation}
Taking $A = 4$ one recovers \cite[eq.~(3.9ii)]{wetterich1988cosmologies}. For $Y > 2/n$ the asymptotic behavior is $A \dot\sigma/n M = 2 H/n Y$ and the ratio $\lambda/\rho_\sigma$ approaches $\frac{1}{2} \, n \, Y - 1$. For $A = n = 4$ this corresponds to the solution \cite[eq.~(3.9i)]{wetterich1988cosmologies}. Comparing these cosmologies with the criteria formulated in \cite[sec.~5]{wetterich1988cosmologies}, we find that the second condition \cite[eq.~(5.2)]{wetterich1988cosmologies} is violated for $Y > \frac{2}{3}$. Helium synthesis and the background radiation would be unacceptably altered for $Y > \frac{1}{2}$. We therefore concentrate on the case $Y < \frac{1}{2}$ which has the standard asymptotic evolution law $H = (2/n) \, t^{-1}$, $\rho \sim t^{-2}$, $\varphi = \text{const}$. A realistic overall cosmological evolution with asymptotically vanishing cosmological ``constant'' emerges provided
\begin{equation}
	A
	> \sqrt{8 (3 + 2 \omega)}.
\end{equation}
It may be surprising that realistic cosmologies are obtained even for $A$ smaller than four ($\omega$ must be negative in this case). Although the potential $V_0$ increases $\sim \chi^{4-A}$ (compare \labelcref{eqn:natural potential}) the field $\chi$ is nevertheless driven to infinity! Due to the coupling to gravity the driving force for $\chi$ is proportional to $4 V_0/\chi - \partial V_0/\partial \chi$ instead of the standard behavior depending only on the derivative of $V_0$. For $0 < A < 4$ the dilatation symmetric solution at $\chi_0 = 0$ is unstable. Instead of approaching the minimum of $V_0$ the field $\chi$ moves upwards in this potential. However, $V_0$ increases slower than $\chi^4$. As a result the ratio $\tilde{R}/\chi^2$ goes to zero and spacetime approaches Minkowski space asymptotically. Indeed, there is no difference between $A$ greater or smaller than four in the Weyl-scaled version. For all positive $A$ the potential $W(\sigma)$ decreases $\sim \exp(-\sigma \, A/M)$ and the cosmon condition $\partial W/\partial \sigma = W
= 0$ is asymptotically fulfilled for $\sigma \to \infty$.

For $Y < \frac{1}{2}$ the main difference between the cosmology discussed in this section and the standard hot big bang evolution is the contribution of the coherent motion of the $\sigma$ field to the total energy density according to \labelcref{eqn:sigma-motion contribution}. This influences the time scale during nucleosynthesis. Applying criterion (v) of \cite[sec.~5]{wetterich1988cosmologies}, this implies an upper bound on $Y$ of
\begin{equation}\label{eqn:y bound}
	Y
	\lesssim \num{0.1}.
\end{equation}
This can be fulfilled even for small values of $A$, provided $\omega$ is near the critical value $\omega_c = -\frac{3}{2}$. We have no independent information on $\omega$ and a small value for $\omega - \omega_c$ may not be unnatural. We recall that for $\omega = \omega_c$ the field $\sigma$ ceases to be a propagating degree of freedom. Also for $\omega = \omega_c$ dilatation symmetry becomes a local instead of a global symmetry. The theory therefore has particular properties for $\omega \to \omega_c$ and a value $\omega$ near $\omega_c$ must not be a ``fine tuning'' of parameters.

Bringing things together we have found a realistic cosmology where Newton's constant decreases with time. It vanishes asymptotically as $\chi$ goes to infinity. In this sense the weakness of gravitational interactions is not intrinsic but rather a consequence of the evolution of the universe. Nevertheless the ratio $\tilde\varphi/\chi$ should reach a (small) constant value asymptotically. This ratio is supposed to be an intrinsic property of the theory. In this respect the model resembles standard cosmology rather than Dirac's hypothesis. Could there be observable consequences of this scenario? Let us first estimate the mass of a ``cosmon'' excitation $\sigma - \sigma_0$ with $\sigma_0$ the coherent background field with cosmological evolution \labelcref{eqn:cosmoglogical evolution}:
\begin{alignedeqn}
	m_\sigma^2
	&= \frac{1}{4 (3 + 2 \omega)} \, \frac{\partial^2 W}{\partial \sigma^2}(\sigma = \sigma_0)\\
	&= \frac{A^2}{4 (3 + 2 \omega)} \, \frac{W(\sigma_0)}{M^2}.
\end{alignedeqn}
Expressing $W(\sigma_0)$ in terms of today's Hubble parameter $H_0$ one finds that today's cosmon mass is given by $H_0$ independent of all other parameters of the model,
\begin{equation}
	m_\sigma^2
	= \frac{9}{2} \, H_0^2.
\end{equation}
We find a new ``universal'' force with a range given by the size of our observable universe!

For all purposes except cosmology this cosmon is massless. The cosmon coupling to matter (take a nucleus, for example) is of gravitational strength ($\sim 1/M^2$). It depends \cite{peccei1987adjusting} on the expectation value of the anomaly $\tensor{\vartheta}{_\mu^\mu}$ in a nucleus\footnote{The formulae in ref.~\cite{peccei1987adjusting} correspond to $\omega = \frac{1}{8}$, $h$ free. They are related to this version ($\omega$ free, $h = 1$) by a rescaling of $\chi$, resulting in the identification $4 (3 + 2 \omega) = (1 + 12 h)/h$. Derivatives and metric in ref.~\cite{peccei1987adjusting} correspond to \cite[sec.~2]{wetterich1988cosmologies}, not to the Weyl-scaled version of \cite[sec.~4]{wetterich1988cosmologies}, which we use in this section.}. In addition there are possible contributions from spatial gradients of fields in a nucleus. We will not attempt in this paper to estimate the cosmon charge $Q_\text{N}$ for the model considered in this section. We only note that as long as $Q_\text{N}$ is proportional to the mass of the nucleus, $M_\text{N}$, one would simply have an additional long range attractive force adding to gravity. Its only consequence would be a difference between the value of Newton's constant observed in our solar system (or galaxy) and the one relevant for nucleosynthesis. The first counts both contributions from the cosmon and the graviton and is therefore larger than the purely gravitational constant $G_\text{N} = M_\text{P}^{-2}$. As a result the Planck mass $M_\text{P}$ could be somewhat higher than commonly estimated. The effect on nucleosynthesis goes in the opposite direction than the effect from $\rho_\sigma$ \labelcref{eqn:sigma-motion contribution} and the bound \labelcref{eqn:y bound} could increase. Deviations from $Q_\text{N} \sim M_\text{N}$ are expected \cite{peccei1987adjusting} to be proportional to baryon number in leading order. They may give rise to a baryon number dependence of the combined graviton plus cosmon force, which does not depend on distance for the model of this section. Experiment tells that such a baryon-number-dependent contribution must be small \cite{pekar1922beitrage,fischbach1986reanalysis,de1986weaker,stubbs1987search,thieberger1987search,boynton1987search,adelberger1987new,niebauer1987galilean}.

The cosmon coupling to matter could also have effects on cosmology by inducing a non-vanishing value $q^\sigma$ in the field equations \cite[eq.~(4.18)]{wetterich1988cosmologies}. This may be particularly important for the matter dominated epoch. Since in our model $\dot\sigma$ does not vanish we would predict a deviation from energy and momentum conservation according to \cite[eq.~(4.18)]{wetterich1988cosmologies}. This would lead for $n = 3$ to an asymptotic behavior $H = \eta t^{-1}$, $\eta \neq \frac{2}{3}$ as discussed in ref.~\cite{reuter1987time}. One also should estimate possible dissipative effects from the decay of the coherent $\sigma$ motion. They could modify the contribution of $\rho_\sigma$ to the total energy density and therefore alter \labelcref{eqn:y bound}. At first sight, however, such effects seem to be very small.

In any case, we should not forget that our model (characterized by \labelcref{eqn:const varphi potential} and $\varphi = \text{const}$) is at best an approximation. It is conceivable that the ratio $\tilde\varphi/\chi$ undergoes a very slow change even for the asymptotic behavior, resulting in a tiny value of $\sigma$ for the discussion of \cite[sec.~5]{wetterich1988cosmologies}. Even for $\varphi = \text{const}$ and a potential $V_0$ fulfilling \labelcref{eqn:cosmological constant rg equation} we expect the anomalous dimension $A$ to depend on the renormalized coupling constants of the theory. This may induce a weak $\chi$-dependence of $A$ -- typically $A = A_0 + A_1 \, \ln(\chi/m)$. For $A$ depending on $\sigma$ one obtains
\begin{equation}
	W_0(\sigma)
	= \bar{W} \, \exp\biggl(-\frac{1}{M} \int_{\bar\sigma}^\sigma A(\sigma) \, \dif \sigma\biggr).
\end{equation}
It is well conceivable that the $\sigma$-dependence of $A$ leads to cosmologies where both $\rho_\sigma$ and $\chi$ decrease faster than $t^{-2}$ so that all effects form $\sigma$ become asymptotically negligible (at least for $q^\sigma = 0$). As an example, consider a potential which can be approximated for $\sigma > -\bar\sigma$ by
\begin{equation}
	W_0(\sigma)
	= \tilde{W} \biggl(\frac{\sigma + \bar\sigma}{M}\biggr)^{-2 \varepsilon} \exp\biggl\{-\bar{A} \biggl(\frac{\sigma + \bar\sigma}{M}\biggr)^{1 + \varepsilon}\biggr\}.
\end{equation}
This leads to an asymptotic solution $H = (2/n) \, t^{-1}$, $\rho \sim t^{-2}$ with
\begin{alignedeqn}
	&\sigma(t)
	= M \biggl(\frac{2}{\bar{A}} \ln \frac{t}{\bar{t}}\biggr)^\frac{1}{1 + \varepsilon} - \bar\sigma,\\
	&\bar{t}^2
	= \frac{8 (3 + 2 \omega) (6/n - 1)}{(1 + \varepsilon)^2 \bar{A}^2} \, \frac{M^2}{\tilde{W}}.
\end{alignedeqn}
Both $\dot\sigma^2$ and $W$ (and therefore $\rho_\sigma$ and $\lambda$) decrease asymptotically like $t^{-2} \, (\ln t)^{-2 \varepsilon/(1 + \varepsilon)}$ and become negligible for $\varepsilon > 0$.

\section{Conclusions}
\label{sec:conclusions}

We have found realistic cosmologies for models where Newton's ``constant'' is a dynamical degree of freedom and can therefore evolve with time. The models we have considered are quite different from Brans-Dicke cosmology due to the existence of a non-trivial effective potential. It is crucial for realistic late cosmology that the ratio between the scales of weak and strong interactions and the dynamical Planck mass, $\tilde\varphi/\chi$, approaches asymptotically a constant (or almost so). Not only the Planck mass ($\sim \chi$) but also the scales of weak and strong interactions ($\sim \tilde\varphi$) should correspond to dynamical degrees of freedom. These scales may also change during the evolution of the universe. Two general types of cosmologies are possible in this context. Either $\chi$ approaches asymptotically a constant value $\chi_0$ and similar for $\tilde\varphi$. Realistic late cosmology is then expected to be of the standard type. Or $\chi$ goes asymptotically to infinity. Depending on the specific model the cosmology can be of the standard type, but interesting modifications, for example for the critical energy density of matter, the age of the universe, or the static behavior of coupling constants, are also possible.

In our models the expectation value of $\chi$ is identified with the scale of spontaneous breaking of dilatation symmetry. The Fermi scale $\tilde\varphi$ and the scale of strong interactions $\Lambda_\text{QCD}$ should asymptotically be proportional to $\chi$. They should therefore not correspond to intrinsic scales of the theory if cosmology is of the type where $\chi$ still evolves today and goes asymptotically to infinity. In this case $\Lambda_\text{QCD}$ and $\tilde\varphi$ should also be a consequence of spontaneous dilatation symmetry breaking. This is realized if the renormalized dimensionless couplings of the scale-invariant version of the standard model, when evaluated at the short distance scale $\chi$, are either independent of $\chi$ or depend only weakly on $\chi$. In the first case dilatation symmetry has no anomaly and the fundamental theory should be finite. (This could be the case for superstrings.) For the second possibility the running of the short distance couplings generates a dilatation anomaly. Scale transformations are not a quantum symmetry and the running of dimensionless couplings introduces an intrinsic scale $m$ in the theory. If the dependence of the short distance couplings on $\chi$ is weak, the intrinsic scale $m$ is much smaller than the spontaneous scale $\chi$. We discussed models where $m$ is even much smaller than $\Lambda_\text{QCD}$ and $\tilde\varphi$ so that $\tilde\varphi/\chi$ is essentially unaffected by the existence of an intrinsic scale and dilatation symmetry is a good approximation for the low energy standard model. There is actually no contradiction between the observed running of the strong coupling constant and the absence (or small role) of dilatation anomalies. The dilatation anomaly is connected to the running of the fundamental coupling constants of the short distance theory and not to the evolution of the effective low energy theory. We also have studied models where strong and/or weak interactions are characterized by an intrinsic scale $m \sim \Lambda_\text{QCD}$ or $m \sim \tilde\varphi$. The language of spontaneously broken dilatation symmetry is still adapted ($m \ll \chi$) and cosmology can be characterized by properties of the dilatation anomaly. In such models the asymptotic constant ratio $\tilde\varphi/\chi$ must follow as a consequence of $\chi$ approaching a constant $\chi_0$.

In this paper we only describe late cosmology. Very early cosmology may be quite different from the asymptotic behavior of $\tilde\varphi/\chi$. For example, $\chi$ may initially have been of the same order as $\tilde\varphi$. This would have important consequences for early cosmology since gravitational interactions would have had the same strength as weak interactions! It is not clear for such scenarios if the temperature was ever high enough to restore $SU(2) \times U(1)$ symmetry. There was possibly no weak phase transition in early cosmology. (In the Weyl-scaled formulation $\varphi$ decreases from a value $\sim M_\text{P}$ to today's scale in this case.)

What about the physics associated with the vanishing of the cosmological constant? In general, a theory has two different systems of mass scales. First, there are intrinsic scales. These are generated by the running of fundamental dimensionless coupling constants and connected to the dilatation anomaly. More generally, if a model has intrinsic mass parameters (like a term $\mu_\varphi^2 \, \tilde\varphi^2$ in the Higgs potential) we may formally include such explicit scale-breaking effects in the anomaly. We denote the largest intrinsic physical mass scale by $m$. Second, we have ``sliding'' scales. These correspond to expectation values of scalar operators. Their value is determined dynamically and may evolve with time. We denote the heaviest sliding scale by $M$. Typically, today's value of $M$ should be in the vicinity of the Planck mass $M_\text{P}$. The two systems of mass scales can move against each other. This corresponds to the degree of freedom of a (pseudo-)dilaton. We call this excitation a cosmon if its dynamics leads to a vanishing cosmological constant. Depending on the ratio $m/M$ we distinguish four different scenarios for the cosmological constant.
\begin{enumerate}[label=(\Alph*),wide,labelwidth=!,labelindent=0pt,nolistsep]
	\item Dilatation symmetry has no anomalies. No intrinsic mass scale appears in the theory ($m = 0$). There is a massless Goldstone boson $\sigma \sim \ln \chi$ (unless its kinetic term vanishes). The vanishing of the cosmological constant is related to dilatation symmetry only through the specific form of the effective potential. This must possess a minimum for non-zero VEVs of some scalars or, equivalently, a flat direction. New physics could only arise if the derivative couplings of the Goldstone boson would lead to an appreciable coupling to matter. This would also influence cosmology for the matter dominated epoch ($q^\sigma \neq 0$).
	
	\item The intrinsic scale $m$ is much smaller than the observed scales of weak and strong interactions. Integrating out all degrees of freedom of the standard model leads to an effective theory for gravity and the field $\chi$ which is characterized by an effective potential $V_0(\chi)$. If the intrinsic mass scale $m$ arises only through the running of fundamental dimensionless coupling constants, the dilatation anomaly is given by the renormalization group equation for $V_0$, $\tensor{\tilde\vartheta}{_\mu^\mu} = \mu \, \partial V_0/\partial \mu$. We assume that this renormalization group equation is determined by a non-vanishing anomalous dimension $\mu \, \partial V_0/\partial \mu = A \, V_0$. This implies a specific form of the potential for $\sigma$, $W(\sigma) \sim \exp(-A \sigma/M)$. As a consequence, the sliding scales ($\chi$, $\tilde\varphi$) still move today compared to $m$ and the ratio $\chi/m$ goes asymptotically to infinity. This leads to cosmologies with a non-trivial time evolution of the cosmological constant which vanishes asymptotically. The kinetic energy of the $\sigma$ field can contribute a fraction $\rho_\sigma$ to the total energy density of the late universe. The cosmon $\sigma$ mediates a new long range force with at most gravitational strength. Its mass is today given by the Hubble parameter, $m_\sigma \sim H_0$.
	
	Depending on the matter couplings of the cosmon this could lead to a composition dependence of the combined gravitational and cosmon force. If the cosmon contributes a substantial amount to the long range force, this would influence the age of the universe. For the matter dominated epoch the energy-momentum tensor of matter would not be conserved ($q^\sigma \neq 0$) and the evolution law could be modified ($H = \eta \, t^{-1}$, $\eta \neq \frac{2}{3}$). Also, if ($\tilde\varphi/\chi$ approaches only asymptotically a constant value, the variation of this quantity today would lead to a time dependence of coupling constants. All these interesting possible effects require, however, a substantial coupling of the pseudo-dilaton $\sigma$ to matter.
	
	\item The intrinsic mass is identified with the scale of strong or weak interactions \cite{peccei1987adjusting}, $m \sim \Lambda_\text{QCD}$ or $m \sim \tilde\varphi$. In this case $\chi$ should approach a static finite value ($\sigma \to \sigma_0$). This requires that the anomalous trace of the energy-momentum tensor must vanish for some value of $\sigma$ ($\tensor{\tilde\vartheta}{_\mu^\mu}(\sigma_0) = 0$). The cosmological constant vanishes if, for $\sigma \to \sigma_0$, the trace of the energy-momentum tensor is purely anomalous ($\tensor{T}{_\mu^\mu}(\sigma_0) = \tensor{\tilde\vartheta}{_\mu^\mu}(\sigma_0)$). The connection of this ``cosmon-condition'' with properties of the fundamental theory is not well understood. The mass of the cosmon is $m_\sigma \sim m^2/M$. Exchange of cosmons leads to an intermediate-range force with gravitational strength and typically a non-trivial composition dependence. Depending on $m$, this force could be observable. The cosmology for this scenario is not yet well studied. It is influenced by the energy stored in the coherent motion of the cosmon. For late cosmology, this depends on the initial conditions after the last (QCD) phase transition and on the matter couplings of the cosmon ($q^\sigma$). Since the anomalous trace of the energy-momentum tensor depends on condensates, and therefore also the value $\sigma_0$ which determines the strength of gravity, the cosmology of phase transitions may be rather complicated.
	
	\item Finally we should mention the possibility that $m \approx M_\text{P}$. No observable long-range or intermediate-range effects survive. All particles except those of the standard model and the graviton have huge masses $\sim M_\text{P}$. Although our treatment of dilatation symmetry and its relation to the cosmological constant remains formally valid, it is questionable that it is helpful for an understanding of the cosmological constant problem. (One word of caution, however, applies to the last two scenarios: It is not completely excluded that intrinsic scales appearing in particle physics are much higher than the one characterizing cosmology.)
	
	It is even conceivable that features of two of our scenarios are realized simultaneously. This can happen if the effective low-energy theory has an additional approximate dilatation-type symmetry, corresponding to a rescaling of $\varphi$ and $g_{\mu\nu}$ in the Weyl-scaled version. (This requires $\Delta V_\chi = 0$.) With respect to such a symmetry $M$ plays the role of an ``intrinsic'' scale and the symmetry is broken explicitly in the gravitational sector. Nevertheless, there may be an additional pseudo-Goldstone boson $\tau$ with a small mass and non-trivial interactions with gravitational strength. Our formalism with $h(\tau) \neq 1$ can be applied to this situation.
	
	Although we know from observation that today's value of the cosmological constant must be tiny, we do not know which one of our scenarios applies. The unknown physics related to this question could well give rise to interesting observable effects. Possible deviations from standard big bang cosmology for the late evolution of the universe could provide important hints about properties of the fundamental theory.
\end{enumerate}

\paragraph{Acknowledgements} 
The author would like to thank R.~D.~Peccei and J.~Solà for the fruitful collaboration on important aspects of this work. He thanks W.~Buchmüller, N.~Dragon, M.~Lüscher and M.Reuter for interesting discussions.

\appendix
\section{Appendix}
\label{sec:appendix}
\allowdisplaybreaks

In order to account for possible dilatation anomalies in the gravitational sector we generalize the gravitational piece of the effective action
\begin{equation}\label{eqn:gravitational action}
	S_\text{G}
	= -\int \dif^4 x \, \tilde{g}^{1/2} \, h \, \chi^2 \, \tilde{R},
\end{equation}
with $h$ a dimensionless function of $\tilde\varphi$ and $\chi$, depending on $\mu$ if there is a dilatation anomaly in the gravitational sector
\begin{equation}
	h
	= h\bigl(\tilde\varphi,\chi;\mu,g_i(\mu)\bigr).
\end{equation}
The action \labelcref{eqn:brans-dicke action} and the field equations of \cite{wetterich1988cosmologies} are recovered for $h = 1$. This modification of the gravitational interactions changes the dilatation current,
\begin{align}\label{eqn:dialtation current appendix}
	J_\text{D}^\mu
	&= 2 \tilde{g}^{1/2} \tilde{g}^{\mu\nu}\biggl\{\biggl(4 \omega + 6 h + 3 \chi \, \frac{\partial h}{\partial \chi}\biggr) \chi \partial_\nu \chi\notag\\
	&\hphantom{{}= 2 \tilde{g}^{1/2} \tilde{g}^{\mu\nu}\biggl\{} + \biggl(1 + \frac{3 \chi^2}{\tilde\varphi} \frac{\partial h}{\partial \tilde\varphi} \tilde\varphi \partial_\nu \tilde\varphi\biggr) \biggr\}\\
	&= 2 g^{1/2} g^{\mu\nu}\biggl\{\biggl(4 \omega + 6 h + \frac{\varphi^2}{M^2} + 3 M \, \frac{\partial h}{\partial \sigma}\biggr) M \partial_\nu \sigma\notag\\
	&\hphantom{{}= 2 g^{1/2} g^{\mu\nu}\biggl\{} + \biggl(1 + \frac{3 M^2}{\varphi} \frac{\partial h}{\partial \varphi} \varphi \partial_\nu \varphi\biggr) \biggr\}.\notag
\end{align}
The gravitational contribution to the dilatation anomaly is
\begin{alignedeqn}
	\tilde\vartheta_\text{G}
	&= -\chi^2 \, \tilde{R} \biggl(\chi \, \frac{\partial h}{\partial \chi} + \tilde\varphi \, \frac{\partial h}{\partial \tilde\varphi}\biggr),\\
	\vartheta_\text{G}
	&= \exp\biggl(-\frac{4 \sigma}{M}\biggr) \tilde\vartheta_\text{G}\\
	&= -M^3 \, \frac{\partial h}{\partial \sigma} \biggl(R + \frac{6}{M} \, \tensor{\sigma}{_;^\mu_\mu} - \frac{6}{M^2} \, \tensor{\sigma}{_;^\mu} \, \tensor{\sigma}{_;_\mu}\biggr).
\end{alignedeqn}
(One may remove a total divergence from $\vartheta_\text{G}$ and add a corresponding piece to $J_\text{D}^\mu$). For $h$ independent of $\mu$, $h = h(\tilde\varphi/\chi)$, one has $\tilde\vartheta_\text{G} = \vartheta_\text{G} = 0$. The equation of motion for $\sigma$ can be derived from the dilatation current \labelcref{eqn:dialtation current appendix} in the Weyl-scaled version (assuming $\partial_\mu \varphi = 0$, $q^\sigma = 0$),
\begin{alignedeqn}\label{eqn:sigma eom}
	&\biggl(8 \omega + 12 h + 12 M\, \frac{\partial h}{\partial \sigma} + \frac{2 \varphi_0^2}{M^2}\biggr) \tensor{\sigma}{_;^\mu_\mu}\\
	&\hphantom{{}=}+ 6 \biggl(\frac{\partial h}{\partial \sigma} + M\, \frac{\partial^2 h}{\partial \sigma^2}\biggr) \sigma_{;\mu} \, \tensor{\sigma}{_;^\mu}\\
	&= -\frac{\partial W}{\partial \sigma} - M^2 \, \frac{\partial h}{\partial \sigma} \, R.
\end{alignedeqn}
Here we define $W(\sigma) = W(\sigma,\varphi_0)$ and similar for $h(\sigma)$.

The modified gravitational equation from \labelcref{eqn:gravitational action} is
\begin{align}
	&\tilde{R}_{\mu\nu} - \frac{1}{2} \tilde{R} \, \tilde{g}_{\mu\nu}\\
	&= \frac{1}{2 h \chi^2} \bigl\{(2 h \chi^2)_{;\mu\nu} - \tensor{(2 h \chi^2)}{_;^\rho_\rho} \, \tilde{g}_{\mu\nu} + 8 \omega \chi_{;\mu} \, \chi_{;\nu}\notag\\
	&\hphantom{{}=} - 4 \omega \tensor{\chi}{_;^\rho} \, \chi_{;\rho} \, \tilde{g}_{\mu\nu} + V \tilde{g}_{\mu\nu} + \tilde{T}_{\mu\nu}^\varphi + \tilde{T}_{\mu\nu}^M\bigr\}.\notag
\end{align}
From this we compute the curvature scalar in the Weyl-scaled version
\begin{alignedeqn}
	R
	&= 3 \, \frac{\partial \ln h}{\partial \sigma} \, \tensor{\sigma}{_;^\mu_\mu} + 3 \, \frac{\partial^2 \ln h}{\partial \sigma^2} \,  \sigma_{;\mu} \, \tensor{\sigma}{_;^\mu}\\
	&\cramped{\hphantom{=}-\frac{1}{2 h M^2} \biggl[4 W - \biggl(8 \omega + 12 h + \frac{2 \varphi^2}{M^2}\biggr)\tensor{\sigma}{_;^\mu} \, \tensor{\sigma}{_;_\mu}}\\
	&\cramped{\hphantom{=-\frac{1}{2 h M^2} \biggl\{}- 2 \varphi_{;\mu} \, \tensor{\varphi}{_;^\mu} - \frac{4 \varphi}{M} \varphi_{;\mu} \, \tensor{\sigma}{_;^\mu} + T_\mu^{M\mu}\biggr].}
\end{alignedeqn}
Insertion into \labelcref{eqn:sigma eom} gives (for $\varphi$ constant)
\begin{align}
	&\begin{aligned}\label{eqn:f+g}
		f(\sigma)& \, \tensor{\sigma}{_;^\mu_\mu} + g(\sigma) \, \tensor{\sigma}{_;^\mu} \, \tensor{\sigma}{_;_\mu}
		= -\frac{\partial W}{\partial \sigma}\\
		&+ 2 \, \frac{\partial \ln h}{\partial \sigma} \, W + \frac{1}{2} \, \frac{\partial \ln h}{\partial \sigma} \, T_\mu^{M \mu}
		= \frac{1}{M} \, \vartheta(\sigma),\\
	\end{aligned}\\
	&\begin{aligned}
		f(\sigma)
		&= 8 \omega + 12 h + \frac{2 \varphi_0^2}{M^2}\\
		&\hphantom{{}=} + 12 M \, \frac{\partial h}{\partial \sigma} + 3 M^2 \, \frac{\partial h}{\partial \sigma} \frac{\partial \ln h}{\partial \sigma},
	\end{aligned}\\
	&\begin{aligned}
		g(\sigma)
		&= 12 \, \frac{\partial h}{\partial \sigma} + 6 M \, \frac{\partial^2 h}{\partial \sigma^2} + \biggl(4 \omega + \frac{\varphi_0^2}{M^2}\biggr) \frac{\partial \ln h}{\partial \sigma}\\ 
		&\hphantom{{}=} + 3 M^2 \, \frac{\partial h}{\partial \sigma} \, \frac{\partial^2 \ln h}{\partial \sigma^2}.
	\end{aligned}
\end{align}
We note that for $\partial h/\partial \sigma \neq 0$ Newton's ``constant'' still varies in the Weyl-scaled version if $\vartheta(\sigma)$. Also $\dot\sigma \neq 0$ has a contribution from the trace of the energy-momentum tensor of matter $T_\mu^{M\mu}$. Such features could be removed using a different Weyl-scaling and a new variable $\sigma^\prime$. In the new scaling, however, $q^\sigma = 0$ would correspond to non-vanishing $q^{\sigma^\prime}$ and the energy-momentum tensor would no longer be conserved in the matter dominated period for $\dot\sigma^\prime \neq 0$. Although we will not discuss this effect in detail in this paper it may have interesting consequences for cosmology in the matter-dominated period.

\end{multicols}

\begin{multicols}{2}[{\printbibheading[title={References}]}]
	\printbibliography[heading=none]
\end{multicols}

\end{document}